\documentclass
[aps,preprint,tightenlines,superscriptaddress,bibnotes,balancelastpage,secnumarabic,titlepage,10pt,twocolumn]{revtex4}%
\usepackage{amsfonts}
\usepackage{amsmath}
\usepackage{amssymb}
\usepackage{graphicx}
\usepackage{hyperref}%
\providecommand{\U}[1]{\protect\rule{.1in}{.1in}}

\begin{document}
\preprint{UMTG-26}
\title[Gali Ge]{Geons of Galileons}
\author{Thomas L. Curtright}
\affiliation{Department of Physics, University of Miami, Coral Gables, FL 33124-8046, USA}
\author{David B. Fairlie}
\affiliation{Department of Mathematical Sciences, Durham University, Durham, DH1 3LE,
UK\medskip\medskip}

\begin{abstract}
We suggest that galileon theories should have an additional self-coupling of
the fields to the trace of their own energy-momentum tensor. \ We explore the
classical features of one such model, in flat 4D spacetime, with emphasis on
solutions that are scalar analogues of gravitational geons. \ We discuss the
stability of these scalar geons, and some of their possible signatures,
including shock fronts.

\end{abstract}
\volumeyear{year}
\volumenumber{number}
\issuenumber{number}
\eid{identifier}
\startpage{1}
\endpage{ }
\maketitle

Galileon theories are a class of models for new scalar fields whose
Lagrangians involve multilinears of first and second derivatives, but whose
nonlinear field equations are nonetheless still only second order. \ They may
be important for the description of large-scale features in astrophysics as
well as for elementary particle theory \cite{d,F}. \ Hierarchies of such
Lagrangians giving rise to such field equations were first discussed
mathematically in \cite{H,FG,FG2,FGM}. \ The simplest example involves a
single scalar field. \ 

This galileon field is usually coupled to all \emph{other} matter through the
trace of the energy-momentum tensor, $\Theta^{\text{(matter)}}$, and is thus
gravitation-like by virtue of the similarity between this universal coupling
and that of the metric $g_{\mu\nu}$ to $\Theta_{\mu\nu}^{\text{(matter)}}$ in
general relativity. \ Indeed, some galileon models have been obtained from
limits of higher dimensional gravitation theories \cite{DGP}.

But \emph{surely}, in a self-consistent theory, for the coupling to be truly
universal, the galileon should also be coupled to its own energy-momentum
trace, even in the flat spacetime limit. \ Some consequences of this
additional self-coupling are considered in this paper.

The action for the lowest non-trivial member of the galileon hierarchy can be
written in various ways upon integrating by parts. \ Perhaps the most compact
and memorable of these is%
\begin{equation}
A_{2}=\tfrac{1}{2}\int\phi_{\alpha}\phi_{\alpha}\phi_{\beta\beta}~d^{n}x\ .
\label{A2}%
\end{equation}
where $\phi$ is the scalar galileon field, $\phi_{\alpha}=\partial\phi\left(
x\right)  /\partial x^{\alpha}$, etc., and where repeated indices are summed
using the Lorentz metric $\delta_{\mu\nu}=\mathrm{diag}\left(  1,-1,-1,\cdots
\right)  $. \ 

It is straightforward to include in $A_{2}$ a covariant coupling to a
background spacetime metric and hence to deduce a symmetric energy-momentum
tensor. \ In the flat-space limit, the result is%
\begin{equation}
\Theta_{\mu\nu}^{\left(  2\right)  }=\phi_{\mu}\phi_{\nu}\phi_{\alpha\alpha
}-\phi_{\alpha}\phi_{\alpha\nu}\phi_{\mu}-\phi_{\alpha}\phi_{\alpha\mu}%
\phi_{\nu}+\delta_{\mu\nu}\phi_{\alpha}\phi_{\beta}\phi_{\alpha\beta}\ .
\label{EnergyMomentum2}%
\end{equation}
This is seen to be conserved,%
\begin{equation}
\partial_{\mu}\Theta_{\mu\nu}^{\left(  2\right)  }=\phi_{\nu}~\mathcal{E}%
_{2}\left[  \phi\right]  \ ,
\end{equation}
upon using the field equation that follows from locally extremizing $A_{2}$,
$0=\delta A_{2}/\delta\phi=-\mathcal{E}_{2}\left[  \phi\right]  $, where%
\begin{equation}
\mathcal{E}_{2}\left[  \phi\right]  \equiv\phi_{\alpha\alpha}\phi_{\beta\beta
}-\phi_{\alpha\beta}\phi_{\alpha\beta}\ . \label{E2}%
\end{equation}

An interesting wrinkle now appears: $\ \Theta_{\mu\nu}^{\left(  2\right)  }$
is not traceless. \ Consequently, the usual form of the scale current,
$x_{\alpha}\Theta_{\alpha\mu}^{\left(  2\right)  }$, is not conserved
\cite{J}. \ On the other hand, the action (\ref{A2}) is homogeneous in $\phi$
and its derivatives, and is clearly invariant under the scale transformations
$x\rightarrow sx$ and $\phi\left(  x\right)  \rightarrow s^{\left(
4-n\right)  /3}\phi\left(  sx\right)  $. \ Hence the corresponding Noether
current must be conserved. \ This current is easily found, especially for
$n=4$, so let us restrict our attention to four spacetime dimensions in the
following. \ 

In that case the trace is obviously a total divergence:%
\begin{equation}
\Theta^{\left(  2\right)  }\equiv\delta_{\mu\nu}\Theta_{\mu\nu}^{\left(
2\right)  }=\partial_{\alpha}\left(  \phi_{\alpha}\phi_{\beta}\phi_{\beta
}\right)  \ .
\end{equation}
That is to say, for $n=4$ the virial is the trilinear $V_{\alpha}=\phi
_{\alpha}\phi_{\beta}\phi_{\beta}$. \ So a conserved scale current is given by
the combination,%
\begin{equation}
S_{\mu}=x_{\alpha}\Theta_{\alpha\mu}^{\left(  2\right)  }-\phi_{\alpha}%
\phi_{\alpha}\phi_{\mu}\ .
\end{equation}
Interestingly, this virial is not a divergence modulo a conserved current, so
this model is \emph{not} conformally invariant despite being scale invariant.
\ Be that as it may, it is not our principal concern here.

Our interest here is that the nonzero trace suggests an additional interaction
where $\phi$ couples directly to its own $\Theta^{(2)}$. \ This is similar to
coupling a conventional \emph{massive} scalar to the trace of its own
energy-momentum tensor \cite{FN}. \ In that previously considered example,
however, the consistent coupling of the field to its trace required an
iteration to all orders in the coupling. \ Upon summing the iteration and
making a field redefinition, the Nambu-Goldstone model emerged. \ But, for the
simplest galileon model in four spacetime dimensions, (\ref{A2}), a consistent
coupling of field and trace is much easier to implement. \ \emph{No iteration
is required.} \ The first-order coupling alone is consistent, after
integrating by parts and ignoring boundary contributions, so that \cite{N0}%
\begin{equation}
-\tfrac{1}{4}\int\phi~\partial_{\alpha}\left(  \phi_{\alpha}\phi_{\beta}%
\phi_{\beta}\right)  ~d^{4}x=\tfrac{1}{4}\int\phi_{\alpha}\phi_{\alpha}%
\phi_{\beta}\phi_{\beta}~d^{4}x\ . \label{TraceTerm}%
\end{equation}
(Similar quadrilinear terms have appeared previously in \cite{DEV,DDE}, only
multiplied there by scalar curvature $R$ so that they would drop out in the
flat spacetime limit that we consider.) \ Consistency follows because
(\ref{TraceTerm}) gives an additional contribution to the energy-momentum
tensor which is \emph{traceless}, in 4D spacetime:%
\begin{equation}
\Theta_{\mu\nu}^{(3)}=\phi_{\mu}\phi_{\nu}\phi_{\alpha}\phi_{\alpha}-\tfrac
{1}{4}\delta_{\mu\nu}\phi_{\alpha}\phi_{\alpha}\phi_{\beta}\phi_{\beta
}\ ,\ \ \ \Theta^{(3)}=0\ .
\end{equation}
Of course, coupling $\phi$ to its own trace may impact the Vainstein mechanism
\cite{V} by changing the effective coupling of $\Theta^{\text{(matter)}}$ to
both backgrounds and fluctuations in $\phi$. \ We leave this as an exercise
for the reader.

Based on these elementary observations, we consider a model with action%
\begin{equation}
A=\int\left(  \tfrac{1}{2}\phi_{\alpha}\phi_{\alpha}-\tfrac{1}{2}\lambda
\phi_{\alpha}\phi_{\alpha}\phi_{\beta\beta}-\tfrac{1}{4}\kappa\phi_{\alpha
}\phi_{\alpha}\phi_{\beta}\phi_{\beta}\right)  ~d^{4}x\ , \label{A}%
\end{equation}
where for the Lagrangian $L$ we take a mixture of three terms: \ the standard
bilinear, the trilinear galileon, and its corresponding quadrilinear
trace-coupling. \ The quadrilinear is reminiscent of the Skyrme term in
nonlinear $\sigma$ models \cite{S} although here the topology would appear to
be always trivial. \ 

The second and third terms in $A$ are logically connected, as we have
indicated. \ But why include in $A$\ the standard bilinear term? \ The reasons
for including this term are to soften the behavior of solutions at large
distances, as will be evident below, and also to satisfy Derrick's criterion
for classical stability under the rescaling of $x$. \ Without the bilinear
term in $L$\ the energy within a spatial volume would be neutrally stable
under a uniform rescaling of $x$, and therefore able to disperse \cite{D,E}.

Similarly, for positive $\kappa$, the last term in $A$ ensures the energy
density of static solutions is always bounded below under a rescaling of the
field $\phi$, a feature that would not be true if $\kappa=0$ but $\lambda
\neq0$. \ So, we only consider $\kappa>0$ in the following. \ But before
discussing the complete $\Theta_{\mu\nu}$ for the model, we note that we
did\emph{\ not} include in $A$ a term coupling $\phi$ to the trace of the
energy-momentum due to the standard bilinear term, namely, $\int\phi
\Theta^{(1)}d^{4}x$, where%
\begin{equation}
\Theta_{\mu\nu}^{(1)}=\phi_{\mu}\phi_{\nu}-\tfrac{1}{2}\delta_{\mu\nu}%
\phi_{\alpha}\phi_{\alpha}\ ,\ \ \ \Theta^{(1)}=-\phi_{\alpha}\phi_{\alpha}\ .
\end{equation}
We have omitted such an additional term in $A$ solely as a matter of taste,
thereby ensuring that $L$ is invariant under constant shifts of the field.
\ Among other things, this greatly simplifies the task of finding solutions to
the equations of motion.

The field equation of motion for the model is $0=\delta A/\delta
\phi=-\mathcal{E}\left[  \phi\right]  $, where
\begin{equation}
\mathcal{E}\left[  \phi\right]  \equiv\phi_{\alpha\alpha}-\lambda\left(
\phi_{\alpha\alpha}\phi_{\beta\beta}-\phi_{\alpha\beta}\phi_{\alpha\beta
}\right)  -\kappa\left(  \phi_{\alpha}\phi_{\beta}\phi_{\beta}\right)
_{\alpha}\ .
\end{equation}
As expected, this field equation is second-order, albeit nonlinear. \ Also
note, under a rescaling of both $x$ and $\phi$, nonzero parameters $\lambda$
and $\kappa$ can be scaled out of the equation. \ Define%
\begin{equation}
\phi\left(  x\right)  =\frac{\lambda}{\kappa}~\psi\left(  \sqrt{\frac{\kappa
}{\lambda^{2}}}x\right)  \ . \label{PhiRescaled}%
\end{equation}
Then the field equation for $\psi\left(  z\right)  $ becomes%
\begin{equation}
\psi_{\alpha\alpha}-\left(  \psi_{\alpha\alpha}\psi_{\beta\beta}-\psi
_{\alpha\beta}\psi_{\alpha\beta}\right)  -\left(  \psi_{\alpha}\psi_{\beta
}\psi_{\beta}\right)  _{\alpha}=0\ , \label{PsiEofM}%
\end{equation}
where $\psi_{\alpha}=\partial\psi\left(  z\right)  /\partial z^{\alpha}$, etc.
\ In effect then, if both $\lambda$ and $\kappa$ do not vanish, it is only
necessary to solve the model's field equation for $\lambda=\kappa=1$.

Though $\mathcal{E}$ is nonlinear, it is nevertheless still true that some
plane waves are exact solutions. \ For \textquotedblleft
light-ray\textquotedblright\ plane waves, $\mathcal{E}\left[  A\exp\left(
ik_{\alpha}x_{\alpha}\right)  \right]  =0$ for constant $A$ and $k_{\alpha}$,
if $k_{\alpha}k_{\alpha}=0$ with $A$ arbitrary. \ In this case, each of the
terms in\ $\mathcal{E}$ vanish separately. \ In fact, light-ray plane waves
are only one among many possible solutions for which\ both $\phi_{\alpha
\alpha}=0$ and $\phi_{\beta}\phi_{\beta}=0$. \ On the other hand, for massive
plane waves, $\mathcal{E}\left[  A\exp\left(  ik_{\alpha}x_{\alpha}\right)
\right]  =0$ if $1/k_{\alpha}k_{\alpha}=-3\kappa A^{2}<0$. \ The latter
\textquotedblleft tachyonic\textquotedblright\ solutions would seem to be less
interesting for real physics.

For static, spherically symmetric solutions,\ $\phi=\phi\left(  r\right)  $,
the field equation of motion becomes%
\begin{equation}
0=\frac{1}{r^{2}}\frac{d}{dr}\left(  r^{2}\left(  \phi^{\prime}+\lambda
\frac{2}{r}\left(  \phi^{\prime}\right)  ^{2}+\kappa\left(  \phi^{\prime
}\right)  ^{3}\right)  \right)  \ .
\end{equation}
where $\phi^{\prime}=d\phi/dr$. This is immediately integrated once to obtain
a cubic equation,
\begin{equation}
r^{2}\phi^{\prime}+2\lambda r\left(  \phi^{\prime}\right)  ^{2}+\kappa
r^{2}\left(  \phi^{\prime}\right)  ^{3}=C\ , \label{SSSFieldEqn}%
\end{equation}
where $C$ is the constant of integration. \ Now, without loss of generality
(cf. (\ref{PhiRescaled}) and (\ref{PsiEofM})) we may choose $\lambda>0$.
\ Then, if $C=0$, either $\phi^{\prime}$ vanishes, or else there are two
solutions that are real only within a finite sphere of radius $r=\sqrt
{\lambda^{2}/\kappa}$. \ These two \textquotedblleft
interior\textquotedblright\ solutions are given\ exactly by%
\begin{equation}
\phi_{\pm}^{\prime}=-\frac{1}{r\kappa}\left(  \lambda\pm\sqrt{\lambda
^{2}-r^{2}\kappa}\right)  \text{ .} \label{C=0Solns}%
\end{equation}
Note that these solutions always have $\phi^{\prime}<0$ within the finite sphere.

Otherwise, if $C\neq0$, then examination of the cubic equation for small and
large $\left\vert \phi^{\prime}\right\vert $ determines the asymptotic
behavior of $\phi^{\prime}$\ for large and small $r$. \ In particular, there
is only one type of asymptotic behavior for large $r$:%
\begin{equation}
\phi^{\prime}\underset{r\rightarrow\infty}{\sim}\frac{C}{r^{2}}\text{ \ \ for
either sign of }C\text{\ .}%
\end{equation}
However, there are two types of behavior for large $\left\vert \phi^{\prime
}\right\vert $, corresponding to small $r$. \ Either%
\begin{equation}
r=\frac{-2\lambda}{\phi^{\prime}\kappa}\left(  1+O\left(  \frac{1}%
{\phi^{\prime}}\right)  \right)
\end{equation}
provided $\phi^{\prime}<0$, but with either sign of $C$; or else
\begin{equation}
r=\frac{1}{\phi^{\prime2}}\left(  \frac{C}{2\lambda}+O\left(  \frac{1}%
{\phi^{\prime}}\right)  \right)
\end{equation}
provided $C>0$, but with either sign of $\phi^{\prime}$. \ The corresponding
real solutions behave as%
\begin{gather}
\phi^{\prime}\underset{r\rightarrow0}{\sim}\frac{-2\lambda}{\kappa r}\text{
\ \ for either sign of }C\text{, or}\\
\phi^{\prime}\underset{r\rightarrow0}{\sim}\pm\sqrt{\frac{C}{2\lambda r}%
}\text{ \ \ provided }C>0\ .
\end{gather}
Comparison of\ the small $r$ behavior to the large $r$ asymptotics shows that
in half these cases the solutions would require zeroes to be real and
continuous for all $r$. \ But such zeroes do not occur. \ Instead, half of the
cases provide real solutions only over a finite interval of $r$, somewhat
similar to the $C=0$ solutions in (\ref{C=0Solns}), but not so easily
expressed, analytically. \ 

The solutions which are real for all $r>0$ boil down to two cases, with small
and large $r$ behavior given by either
\begin{equation}
\phi^{\prime}\underset{r\rightarrow0}{\sim}\sqrt{\frac{C}{2\lambda r}%
}\text{\ \ \ and \ \ }\phi^{\prime}\underset{r\rightarrow\infty}{\sim}\frac
{C}{r^{2}}\text{ \ \ for }C>0\text{,} \label{C>0Asymps}%
\end{equation}
or else%
\begin{equation}
\phi^{\prime}\underset{r\rightarrow0}{\sim}\frac{-2\lambda}{\kappa
r}\text{\ \ \ and \ \ }\phi^{\prime}\underset{r\rightarrow\infty}{\sim}%
\frac{C}{r^{2}}\text{ \ \ for }C<0\text{.} \label{C<0Asymps}%
\end{equation}
From further inspection of the cubic equation to determine the behavior of
$\phi^{\prime}$ for intermediate values of $r$, when $C>0$ it turns out that
$\phi^{\prime}$ is a single-valued, positive function for all $r>0$, joining
smoothly with the asymptotic behaviors given in (\ref{C>0Asymps}).
\ However,\ it also turns out there is an additional complication when $C<0$.
\ In this case there is a critical value $\left(  \kappa^{3/2}/\lambda
^{2}\right)  C_{\text{critical}}=-4\sqrt{3}/27\approx-0.2566$ such that, if
$C\leq C_{\text{critical}}$ then $\phi^{\prime}$ is a single-valued, negative
function for all $r>0$, while if $C_{\text{critical}}<C<0$ then $\phi^{\prime
}$ is triple-valued for an open interval in $r>0$. \ It is not completely
clear to us what physics underlies this multivalued-ness for some negative
$C$. \ But in any case, when $C<0$\ it is also true that $\phi^{\prime}$ joins
smoothly with the asymptotic behaviors given in (\ref{C<0Asymps}). \ All this
is illustrated in Figures 1 and 2, for $\lambda=\kappa=1$.

A test particle coupled by $\phi\Theta^{\text{(matter)}}$ to any of these
galileon field configurations would see an effective potential which is not
$1/r$, for intermediate and small $r$. \ Therefore its orbit would show
deviations from the usual Kepler laws, including precession at variance with
that predicted by conventional general relativity. \ It would be interesting
to search for such effects, say, by considering stars orbiting around the
galactic center.%
\begin{figure}[ptb]%
\centering
\includegraphics[
trim=0.000000in 0.000000in 0.000000in 0.036382in,
height=2.1075in,
width=3.3166in
]%
{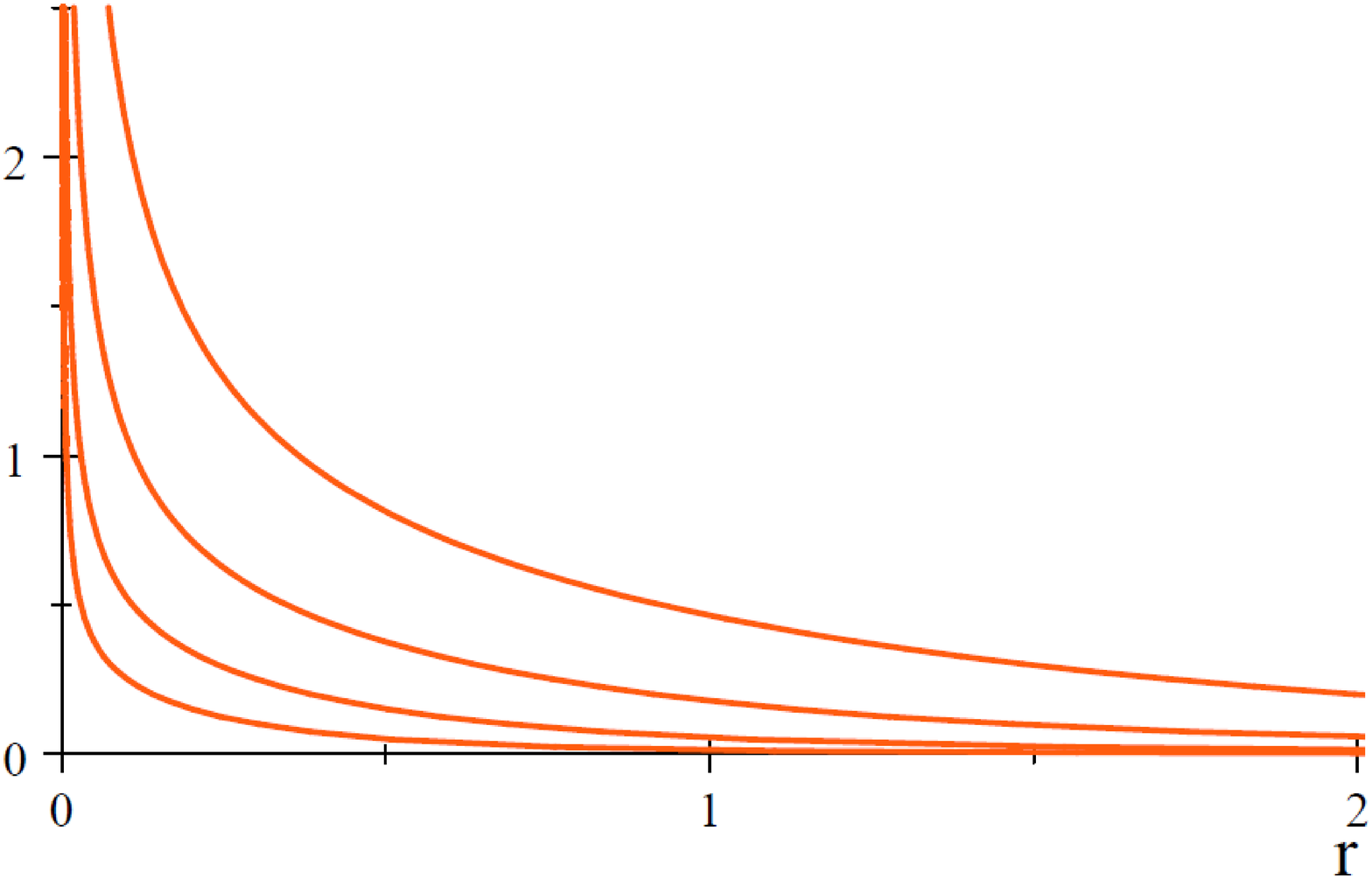}%
\caption{$\psi^{\prime}\left(  r\right)  $ for $C=+1/4^{N}$, with $N=0,1,2,3$
for top to bottom curves, respectively.}%
\end{figure}
\begin{figure}[ptb]%
\centering
\includegraphics[
trim=0.000000in 0.000000in 0.000000in -0.196468in,
height=2.1075in,
width=3.307in
]%
{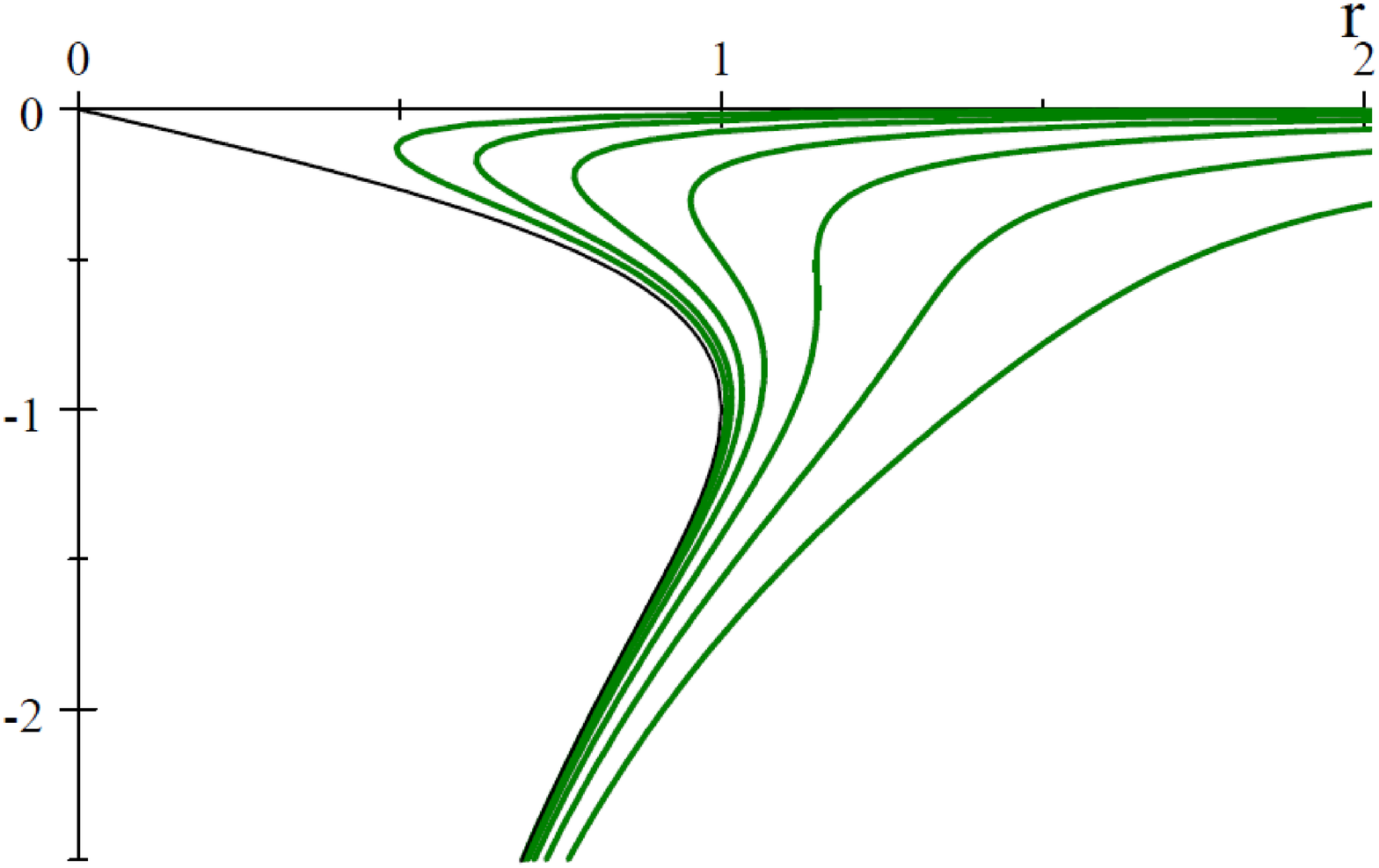}%
\caption{$\psi^{\prime}\left(  r\right)  $ for $C=-1/2^{N}$, with
$N=6,5,4,3,2,1,0$ from left to right, respectively. \ The thin black curve is
a union of the two $C=0$ solutions in (\ref{C=0Solns}).}%
\end{figure}

For the solutions described by (\ref{C>0Asymps}) and (\ref{C<0Asymps}), the
total energy outside any large radius is obviously finite for both $C>0$ and
$C<0$. \ And if $C>0$, the total energy within a small sphere surrounding the
origin is also manifestly finite. \ But if $C<0$ the energy within that same
small sphere could be infinite \emph{unless} there is a cancellation between
the galileon term and the trace interaction term. \ Remarkably, this
cancellation does occur \cite{N1}. \ So both $C>0$ and $C<0$ types of static
solutions for the model have finite total energy. \ 

Complete information about the distribution of energy is provided by the
model's energy-momentum tensor,%
\begin{equation}
\Theta_{\mu\nu}=\Theta_{\mu\nu}^{(1)}-\lambda\Theta_{\mu\nu}^{\left(
2\right)  }-\kappa\Theta_{\mu\nu}^{(3)}\ .
\end{equation}
As expected, this is conserved, given the field equation $\mathcal{E}\left[
\phi\right]  =0$, since%
\begin{equation}
\partial_{\mu}\Theta_{\mu\nu}=\phi_{\nu}\mathcal{E}\left[  \phi\right]  \ .
\end{equation}
The energy density for \emph{static} solutions differs from the canonical
energy density for such solutions (namely, $-L$) by a total spatial divergence
that arises from the galileon term:
\begin{equation}
\Theta_{00}=-\left.  L\right\vert _{\text{static}}-\tfrac{1}{2}\lambda
\overrightarrow{\nabla}\cdot\left(  \left(  \nabla\phi\right)  ^{2}%
\overrightarrow{\nabla}\phi\right)  \ . \label{CvsSymmetric}%
\end{equation}
This divergence will not contribute to the total energy for fields such that
$\lim_{r\rightarrow\infty}\left(  \phi/\ln r\right)  $ exists. \ Assuming that
is the case, Derrick's scaling argument for static, finite energy
solutions\ of the equations of motion \cite{D} shows the energy is just twice
that due to the bilinear $\Theta_{00}^{\left(  1\right)  }$. \ Thus,
\begin{equation}
E=\int\Theta_{00}~d^{3}r=\int\left(  \overrightarrow{\nabla}\phi\right)
^{2}~d^{3}r\ .
\end{equation}

For the spherically symmetric static solutions of (\ref{SSSFieldEqn}), this
becomes an expression of the energy as a function of the parameters and the
constant of integration $C$:
\begin{equation}
E\left[  \lambda,\kappa,C\right]  =4\pi\int_{0}^{\infty}\left(  \phi^{\prime
}\right)  ^{2}~r^{2}dr\ .
\end{equation}
Again without loss of generality, consider $\lambda=\kappa=1$. \ Then for
either $C>0$ or for $C<C_{\text{critical}}<0$ \cite{N2}, change integration
variables from $r$ to $s\equiv\phi^{\prime}$ to find:%
\begin{align}
E\left(  C\gtrless0\right)   &  =I\left(  \left\vert C\right\vert \right)
\mp\left(  \left\vert C\right\vert +\tfrac{1}{2}\pi\right)  \ ,\label{E(C)}\\
I\left(  C>0\right)   &  \equiv\tfrac{1}{2}\int_{0}^{\infty}\frac{P\left(
s,C\right)  ~ds}{\left(  s^{2}+1\right)  ^{4}R\left(  s,C\right)  }\ ,
\label{I(C)}%
\end{align}
where $R\left(  s,C\right)  =\sqrt{s^{4}+s\left(  s^{2}+1\right)  C}$ and
where the numerator of the integrand is an eighth-order polynomial in $s$,
namely,$\ P\left(  s,C\right)  =8s^{8}+12Cs^{7}+\left(  3C^{2}-8\right)
s^{6}+8Cs^{5}+7C^{2}s^{4}-4Cs^{3}+5C^{2}s^{2}+C^{2}$. \ Thus, $I\left(
C\right)  $ is an elliptic integral. \ But rather than express the final
result in terms of standard functions, it suffices here just to plot $E\left(
C\right)  $, in Figure 3. \ Note that $E$ increases monotonically with
$\left\vert C\right\vert $.

For other values of $\lambda$ and $\kappa$ with the constant of integration
$C$ specified as in (\ref{SSSFieldEqn}), the energy of the solution is given
in terms of the function defined by (\ref{E(C)},\ref{I(C)}):
\begin{equation}
E\left[  \lambda,\kappa,C\right]  =\left(  \lambda^{3}/\kappa^{5/2}\right)
~E\left(  \kappa^{3/2}C/\lambda^{2}\right)  \ .
\end{equation}
The energy curves indicate double degeneracy in $E$, for different values of
$\left\vert C\right\vert $, when $E\left[  \lambda,\kappa,C\right]
>\pi\lambda^{3}/\kappa^{5/2}$. \ Also, for a given $\left\vert C\right\vert $
the negative $C$ solutions are \emph{higher} in energy, with $E\left[
\lambda,\kappa,-\left\vert C\right\vert \right]  -E\left[  \lambda
,\kappa,\left\vert C\right\vert \right]  =\pi\lambda^{3}/\kappa^{5/2}%
+2\left\vert C\right\vert \lambda/\kappa$. \ Or at least this is true for all
$\left\vert C\right\vert \geq\left\vert C_{\text{critical}}\right\vert $ in
which case $E\left[  \lambda,\kappa,C\right]  \geq\frac{\lambda^{3}}%
{\kappa^{5/2}}E\left(  \frac{\kappa^{3/2}}{\lambda^{2}}C_{\text{critical}%
}\right)  \approx3.7396~\lambda^{3}/\kappa^{5/2}$ \cite{N2}.
\begin{figure}[ptb]%
\centering
\includegraphics[
trim=0.000000in 0.000000in 0.376326in 0.266826in,
height=2.1926in,
width=3.2943in
]%
{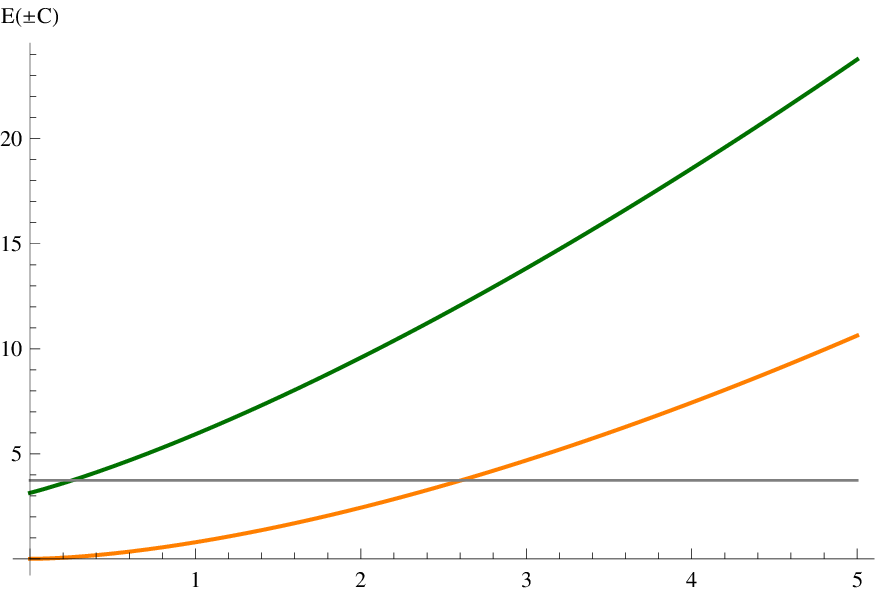}%
\caption{$E\left(  \pm C\right)  $ versus $C\geq0$ as lower/upper curves (the
horizontal line is $E\left(  C_{\text{critical}}\right)  \approx3.7396$).}%
\end{figure}

Finite energy classical solutions of gravity-like theories bring to mind the
\textquotedblleft geons\textquotedblright\ proposed long ago by Wheeler
\cite{W}. \ These were envisioned in their purest form as distributions of
only gravitational energy held together solely by gravitational interaction.
$\ $Combinations of electromagnetic energy and gravity were also considered,
as were systems containing neutrinos. $\ $Wheeler argued that such
configurations would be \emph{relatively}\ stable, if they existed, but would
eventually dissipate due to a variety of both classical and quantum effects,
including\ light-light scattering, as well as production and absorption of
quanta. $\ $While plausible distributions were sketched, and decay rates were
estimated, \emph{exact} classical solutions were not found. $\ $

The same mechanisms would seem to apply to any hypothetical classical galileon
distributions such as those discussed here, the main difference being that
analytic spherically symmetric solutions might still be obtainable even if
conventional gravitational effects were included. \ Perhaps these
gravitational effects would not alter the qualitative features of the static
pure $\phi$ configurations given above. \ Should they really exist, presumably
these galileon geons could also be dissipated by various classical and quantum
effects. \ All this is far beyond our current abilities and the scope of this
paper, of course, but the general ideas suggest some interesting possibilities.

Whatever the cause, if the configuration's energy loss were gradual, as a
first step it might suffice to model the time-dependent system
quasi-statically, as a continuous flow from one static solution to another.
\ That is to say, perhaps a good approximation would be to take $C\left(
t\right)  $, with $\left\vert C\right\vert $ and $E\left(  C\right)  $
decreasing monotonically with time. \ For the positive $C$ case, this would be
more or less uneventful as the whole configuration would just slowly disappear
without any abrupt changes. \ But for the negative $C$ case, as $t$ increased
$C_{\text{critical}}$ would be reached, beyond which the solution would begin
to fold over, exhibiting the multivalued features shown in Figure 2. \ But
this is just the usual picture for the formation of a shock front. \ These
particular galileon shocks would implode, converging towards the origin, as
shown \href{http://server.physics.miami.edu/~curtright/PsiWave.gif}{here}.
\ We believe this is a plausible scenario and a reasonable physical
interpretation of the model's multivalued solutions. \ Moreover, it would seem
to provide a signature for their existence.

As is clear from Figure 2, the shock front would form when $d\phi^{\prime
}/dr=\infty$. \ For the $C<0$ static solutions of (\ref{SSSFieldEqn}) it is
not difficult to determine the locus of such singular points. \ It is given by
the intersection of the solutions, for various $C$, and the curve $\left(
1+3\kappa\phi^{\prime2}\right)  r=4\lambda\phi^{\prime}$. \ As usual for
singular points in the development of a shock, almost certainly there is some
physics missing from the equations. \ Since $\phi^{\prime\prime}$ is large,
the obvious modification would be to include higher derivative terms in the
action, which is tantamount to attempting an ultraviolet completion of the
model. \ This is an open question. \ Perhaps higher terms in the galileon
hierarchy would be natural candidates to be included.

To get a handle on such terms, and for purposes of comparison to the model in
(\ref{A}), consider briefly another model somewhat similar in form, but whose
Lagrangian consists only of terms taken from the galileon hierarchy, without
any coupling to $\Theta$. \ After rescaling the field and coordinates to
achieve a standard form, this alternate model may be defined by%
\begin{gather}
A_{\text{self-dual}}\left[  \psi\right]  =\int\left(  \tfrac{1}{2}\psi
_{\alpha}\psi_{\alpha}-\tfrac{1}{4}\psi_{\alpha}\psi_{\alpha}\psi_{\beta\beta
}\right. \nonumber\\
\left.  +\tfrac{1}{12}\psi_{\alpha}\psi_{\alpha}\left(  \psi_{\beta\beta}%
\psi_{\gamma\gamma}-\psi_{\beta\gamma}\psi_{\beta\gamma}\right)  \right)
d^{4}x\ . \label{ASelfDual}%
\end{gather}
The difference with (\ref{A})\ lies in the last term, which is quadrilinear in
the field, as before, but now has two fields with second derivatives. \ 

As the name suggests, this model is self-dual, in the following sense: \ The
action retains its form under a Legendre transformation \cite{FG2} (also see
\cite{G}) to a new field $\Psi$ and new coordinates $X$, as defined by:%
\begin{equation}
\psi\left(  x\right)  +\Psi\left(  X\right)  =x_{\alpha}X_{\alpha}\ .
\end{equation}
Thus $A_{\text{self-dual}}\left[  \psi\right]  =A_{\text{self-dual}}\left[
\Psi\right]  $, provided integrations by parts give no surface contributions.
\ This identity suggests that there are interesting properties for the
quantized model, such as its ultraviolet behavior, but that is outside the
scope of the present discussion. \ 

Here it suffices to compare the classical physics following from
(\ref{ASelfDual}) with that following from (\ref{A}). \ Upon integrating once
the classical equations of motion for static, spherically symmetric solutions
of the field equations for (\ref{ASelfDual}), the result is again a cubic
equation,%
\begin{equation}
r^{2}\psi^{\prime}+r\left(  \psi^{\prime}\right)  ^{2}+\tfrac{1}{3}\left(
\psi^{\prime}\right)  ^{3}=C\ , \label{SDSSSFieldEqn}%
\end{equation}
but the $\left(  \psi^{\prime}\right)  ^{3}$ term is no longer weighted by
$r^{2}$ as it was in (\ref{SSSFieldEqn}). \ Thus the small and large $r$
behaviors are now given by%
\begin{equation}
\psi^{\prime}\underset{r\rightarrow0}{\sim}\left(  3C\right)  ^{1/3}%
\text{\ \ \ and \ \ }\psi^{\prime}\underset{r\rightarrow\infty}{\sim}\frac
{C}{r^{2}}\text{,}%
\end{equation}
for either sign of the constant of integration, $C$. \ These static solutions
have finite total energy for either sign of $C$, as before, only now
$\psi^{\prime}$ is always bounded. \ Moreover, upon inspection of the behavior
of $\psi^{\prime}$ for intermediate $r$, and various $C$, unlike the previous
model the solutions are now always single-valued for either $C>0$ or $C<0$.
\ Thus there are no multivalued solutions like those shown in Figure 2.
\ However, each of the $C<0$ static solutions does have a single point for
which $d\psi^{\prime}/dr=\infty$, namely, $r=\left(  3\left\vert C\right\vert
\right)  ^{1/3}$. \ So there is still a reason to expect the existence of
shock fronts for quasi-static time-dependent fields in this alternate model.
\ Finally, again for $C<0$, to have $\phi^{\prime}$ real for all $r>0$, it is
necessary to join together \textquotedblleft interior\textquotedblright\ and
\textquotedblleft exterior\textquotedblright\ solutions at $r=\left(
3\left\vert C\right\vert /2\right)  ^{1/3}$.

It remains to investigate the stability of these spherically symmetric
solutions under perturbations, especially to check for the existence of
superluminal modes, along the lines of \cite{GHT}. \ Evidently, superluminal
modes are a possible feature for models of this type.

In conclusion, it would be interesting to search for evidence of geons
containing galileons at all distance scales, including galactic and
sub-galactic, as well as cosmological. \ Perhaps a combination of trace
couplings and various galileon terms, such as those in (\ref{A}) and
(\ref{ASelfDual}), will ultimately lead to a realistic physical model.\medskip

\begin{acknowledgments}
\textit{We thank S Deser and C Zachos for comments on the manuscript. \ This
research was supported by a University of Miami Cooper Fellowship, and by NSF
Award 0855386.}
\end{acknowledgments}


\begin{thebibliography}{99}
\bibitem {N0}Also note that $A_{2}$ follows from coupling $\phi$ to the trace
of the manifestly chargeless tensor $\left(  \partial_{\mu}\partial_{\nu
}-\delta_{\mu\nu}\partial_{\alpha}\partial_{\alpha}\right)  \phi_{\beta}%
\phi_{\beta}$.
\bibitem {N2}The multivalued behavior of any solution for $C_{\text{critical}%
}<C<0$ makes the determination of the total energy ambiguous, at best, for
these cases. \ This is an unresolved issue.
\end{thebibliography}
\end{document}